\documentclass{article}
\usepackage[utf8]{inputenc}
\usepackage{tikz}
\usetikzlibrary{decorations.pathreplacing,arrows,shapes,shapes.geometric}

\title{A Note on the Complexity of Computing the Number of Reachable Vertices in a Digraph}
\author{Michele Borassi}
\date{\today}

\begin{document}

\maketitle

\begin{abstract}
    In this work, we consider the following problem: given a digraph $G=(V,E)$, for each vertex $v$, we want to compute the number of vertices reachable from $v$. In other words, we want to compute the out-degree of each vertex in the transitive closure of $G$. We show that this problem is not solvable in time $\mathcal{O}\left(|E|^{2-\epsilon}\right)$ for any $\epsilon>0$, unless the Strong Exponential Time Hypothesis is false. This result still holds if $G$ is assumed to be acyclic.
\end{abstract}

\section{Introduction}

In this work, we consider the following problem: given a digraph $G=(V,E)$, for each vertex $v$, we want to compute the number of vertices reachable from $v$. An efficient solution of this problem could have many applications: to name a few, there are algorithms that need to compute (or estimate) these values \cite{Borassi2014a}, the number of reachable vertices is used in the definition of other measures, like closeness centrality \cite{Lin1976,Wasserman1994,Olsen2014}, and it can be useful in the analysis of the transitive closure of a graph (indeed, the out-degree of a vertex $v$ in the transitive closure is the number of vertices reachable from $v$). 

Until now, the best algorithms to solve this problem explicitly compute the transitive closure of the input graph, and then output the out-degree of each node. This can be done through fast matrix multiplication \cite{Fischer1971}, in time $\mathcal{O}(N^{2.373})$ \cite{DBLP:conf/stoc/Williams12}, or by performing a Breadth-First Search from each node, in time $\mathcal{O}(MN)$, where $N=|V|$ and $M=|E|$.

However, one might think that if only the number of reachable vertices is needed, then there might be a faster algorithm: in this work, we prove that this is not the case, even if the input graph is acyclic. Indeed, an algorithm running in time $\mathcal{O}(M^{2-\epsilon})$ would falsify the well-known Strong Exponential Time Hypothesis \cite{Impagliazzo2001}: this hypothesis says that, for each $\delta>0$, if $k$ is big enough, the $k$-\textsc{Satisfiability} problem on $n$ variables cannot be solved in time $\mathcal{O}((2-\delta)^{n})$. 
As far as we know, this reduction has never been published, even if several similar reductions are available in the literature \cite{Williams2005,Williams2010,Patrascu2010,Roditty2013,Abboud2014,Abboud2014a,Bringmann2014,Abboud2015,Borassi2015a,Abboud2016}.

\section{The Reduction}

Let us consider an instance of the $k$-\textsc{Satisfiability} problem on $n$ variables, and let us assume that $n=2l$ (if $n$ is odd, we add one variable that does not appear in any clause). Let us divide the variables in two sets $X,Y$, such that $|X|=|Y|=l$. We will name $x_1,\dots,x_l$ the variables in $X$, and $y_1,\dots,y_l$ the variables in $Y$. From this instance of the $k$-\textsc{Satisfiability} problem, let us construct a digraph as follows. We consider the set $V_X$ of all $2^{|X|}$ possible evaluations of the variables in $X$, and the set $V_Y$ of all $2^{|Y|}$ possible evaluations of the variables in $Y$. The set of vertices is $V_X \cup V_Y \cup V_C$, where $V_C$ is the set of clauses.

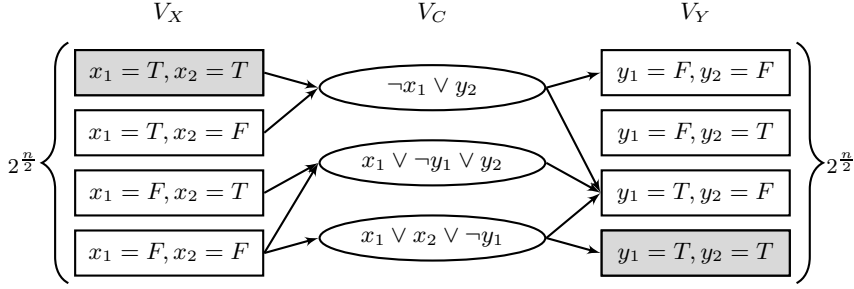
\begin{figure}
\begin{tikzpicture}
\tikzset{every node/.style={rectangle,draw=black,minimum height=.6cm,minimum width=2.5cm,inner sep=0cm,font=\small},
clause/.style={ellipse, minimum width= 3cm},
every path/.style={>=latex',thick}}

\node[draw=none] (x1) at (0,3.2) {$V_X$};
\node[draw=none] (x1) at (3.5,3.2) {$V_C$};
\node[draw=none] (x1) at (7,3.2) {$V_Y$};

\node (x1) at (0,2.4) {$x_1=T,x_2=T$};
\node (x2) at (0,1.6) {$x_1=T,x_2=F$};
\node (x3) at (0,.8) {$x_1=F,x_2=T$};
\node (x4) at (0,0) {$x_1=F,x_2=F$};

\node (y1) at (7,0) {$y_1=T,y_2=T$};
\node (y2) at (7,.8) {$y_1=T,y_2=F$};
\node (y3) at (7,1.6) {$y_1=F,y_2=T$};
\node (y4) at (7,2.4) {$y_1=F,y_2=F$};

\node[clause] (c1) at (3.5,2.2) {$\neg x_1 \vee y_2$};
\node[clause] (c2) at (3.5,.2) {$x_1 \vee x_2 \vee \neg y_1$};
\node[clause] (c3) at (3.5,1.2) {$x_1 \vee \neg y_1 \vee y_2$};

\draw[->] (x1.east) -- (c1.west);
\draw[->] (x2.east) -- (c1.west);
\draw[->] (x3.east) -- (c3.west);
\draw[->] (x4.east) -- (c3.west);
\draw[->] (x4.east) -- (c2.west);

\draw[->] (c2.east) -- (y1.west);
\draw[->] (c2.east) -- (y2.west);

\draw[->] (c1.east) -- (y2.west);
\draw[->][->] (c1.east) -- (y4.west);

\draw[->] (c3.east) -- (y2.west);

\node[fill=gray!30] (x1) at (0,2.4) {$x_1=T,x_2=T$};
\node[fill=gray!30] (y1) at (7,0) {$y_1=T,y_2=T$};

\draw [decorate,decoration={brace,amplitude=10pt},xshift=-4pt,yshift=0pt]
(-1.2,-.4) -- (-1.2,2.8) node [draw=none, fill=none, minimum width=0cm,midway,xshift=-0.6cm] {\footnotesize $2^{\frac{n}{2}}$};

\draw [decorate,decoration={brace,amplitude=10pt, mirror},xshift=4pt,yshift=0pt]
(8.2,-.4) -- (8.2,2.8) node [draw=none, fill=none, minimum width=0cm,midway,xshift=0.6cm] {\footnotesize $2^{\frac{n}{2}}$};
\end{tikzpicture}
\caption{An example of the graph obtained from the formula $(\neg x_1 \vee y_2) \wedge (x_1 \vee \neg y_1 \vee y_2) \wedge (x_1 \vee x_2 \vee \neg y_1)$. The two gray evaluations correspond to a satisfying assignment.} \label{fig:example}
\end{figure}

We add a directed edge from a vertex $v \in V_X$ to a vertex $w \in V_C$ if the evaluation $v$ does not make the clause $w$ true. For instance, if $X=\{x_1,x_2\}$, and $w$ is $x_1 \vee y_2$, the evaluation $(x_1=T, x_2=T)$ is not connected to $w$, because the variable $x_1$ makes the clause $w$ true. Conversely, the evaluation $(x_1=F, x_2=T)$ is connected to $w$, because it does not make the clause $w$ true (note that, in this case, we still can make $w$ true by setting $y_2=T$). Similarly, we add a directed edge from a clause $w \in V_C$ to an evaluation $v$ in $V_Y$ if the evaluation $v$ does not make the clause $w$ true. An example is shown in Figure~\ref{fig:example}.

The formula is satisfiable if and only if we can find an evaluation $v_X \in V_X$ of the variables in $X$ and an evaluation $v_Y \in V_Y$ of the variables in $Y$ such that each clause is either satisfied by $v_X$ or by $v_Y$. By construction, this happens if and only if $v_X$ is not connected to $v_Y$ in the graph constructed (for example, the two gray evaluations in Figure~\ref{fig:example} correspond to a satisfying assignment).

Moreover, the graph constructed has at most $N=|X|+|Y|+|C| \leq 2*2^{\frac{n}{2}}+n^k=\mathcal{O}\left(2^{\frac{n}{2}}\right)$ nodes, and at most $M=|X||C|+|Y||C| \leq 2*2^{\frac{n}{2}}*n^k=\mathcal{O}\left(2^{\frac{n}{2}}n^k\right)$ edges.

This means that, if we can count the number of reachable vertices in time $\mathcal{O}(N^{2-\epsilon})$, then we can also verify if the formula is satisfiable, by checking if all vertices in $V_X$ can reach all vertices in $V_Y$ with no overhead (the number of vertices in $V_Y$ reachable from a vertex $v \in V_X$ can be computed in time $\mathcal{O}(n^k)$ as the total number of vertices reachable from $v$, minus the number of vertices in $V_C$ reachable from $v$). As a consequence, if we have an algorithm that computes the number of reachable vertices in time $\mathcal{O}(M^{2-\epsilon})$ for some $\epsilon$, then we can find an algorithm that solves $k$-\textsc{Satisfiability} in time $\mathcal{O}\left(\left(2^{\frac{n}{2}}n^k\right)^{2-\epsilon}\right)=\mathcal{O}\left(\left(2^{\frac{2-\epsilon}{2}}\right)^n n^{(2-\epsilon)k}\right)=\mathcal{O}\left(\left(2-\delta\right)^n\right)$ for a suitable choice of $\delta$. This falsifies the Strong Exponential Time Hypothesis, and concludes the reduction.

\section*{Acknowledgements}

The author thanks Emanuele Natale for reading carefully and correcting the first version. He also thanks Emanuele Natale and Massimo Cairo for suggesting him to write the short paper.
\bibliographystyle{plain}
\bibliography{library}
\end{document}